\documentclass[fleqn]{2023SCGE}
\usepackage{graphicx}%
\usepackage{subfigure}%
\usepackage{acronym}%
\usepackage{multirow}%
\usepackage{amsmath,amssymb,amsfonts}%

\begin{document}

\ensubject{subject}

\ArticleType{Article}
\Year{2024}
\Month{June}
\Vol{67}
\No{7}
\DOI{https://doi.org/10.1007/s11433-023-2377-7}
\ArtNo{000000}
\ReceiveDate{December 8, 2023}
\AcceptDate{March 15, 2024}
\OnlineDate{June 14, 2024}

\title{Near Real-Time Gravitational Wave Data Analysis of the Massive Black Hole Binary with TianQin}{Near Real-Time Gravitational Wave Data Analysis of the Massive Black Hole Binary with TianQin}

\author[1]{Hong-Yu Chen}{}
\author[1]{Xiang-Yu Lyu}{}
\author[1]{En-Kun Li}{lienk@sysu.edu.cn}
\author[1]{Yi-Ming Hu}{huyiming@sysu.edu.cn}

\AuthorMark{H.-Y. Chen}

\AuthorCitation{H.-Y. Chen, X.-Y. Lyu, E.-K. Li, and Y.-M. Hu}

\address[1]{MOE Key Laboratory of TianQin Mission, %
    TianQin Research Center for Gravitational Physics $\&$ School of Physics and Astronomy, \\%
    Frontiers Science Center for TianQin, %
    Gravitational Wave Research Center of CNSA, %
    Sun Yat-sen University (Zhuhai Campus), %
    Zhuhai, 519082, China}


\abstract{Space-borne gravitational wave (GW) detectors can detect the merger of massive black holes.
The early warning and localization of GW events before merging can be used to inform electromagnetic telescopes and conduct multimessenger observations. 
However, this requires real-time data transmission and analysis capabilities.
The geocentric orbit of the space-borne GW detector TianQin makes it possible to conduct real-time data transmission.
In this study, we develop a search and localization pipeline for massive black hole binaries (MBHBs) with TianQin under both regular and real-time data transmission modes.
We demonstrate that, with real-time data transmission, MBHBs can be accurately localized on the fly.
With the approaching merger, each analysis can be finished in only 40 min.
For an MBHB system at a distance of 1 Gpc, if we receive data every hour, then we can pinpoint its location to within less than 1 deg$^2$ on the final day before the merger.
}%



\maketitle

\acrodef{GW}{gravitational wave}
\acrodef{SBBH}{stellar mass binary black hole}
\acrodef{PTA}{pulsar timing array}
\acrodef{EMRI}{extreme mass ratio inspiral}
\acrodef{DWD}{double white dwarf}
\acrodef{SGWB}{stochastic GW background}
\acrodef{SNR}{signal-to-noise ratio}
\acrodef{LISA}{Laser Interferometer Space Antenna}
\acrodef{EM}{electromagnetic}
\acrodef{TDI}{time delay interferometry}
\acrodef{PSD}{power spectral density}
\acrodef{MCMC}{Markov chain Monte Carlo}
\acrodef{FOV}{field of view}


\begin{multicols}{2}

\section{Introduction}\label{sec1}

The past years have witnessed significant progress in the field of \ac{GW} astronomy, with nearly a hundred \ac{GW} events detected by ground-based \ac{GW} detectors \cite{LIGOScientific:2018mvr,LIGOScientific:2020ibl,LIGOScientific:2021djp}. Recently, some \acp{PTA} have detected $n{\rm Hz}$ \acp{SGWB} \cite{NANOGrav:2023gor,Antoniadis:2023ott,Reardon:2023gzh,Xu:2023wog}.
Meanwhile, \ac{GW} signals in the range of $0.1 ~m{\rm Hz} -- ~ 1 ~{\rm Hz}$ can be detected by space-borne \ac{GW} detectors, such as TianQin \cite{TianQin} and \ac{LISA} \cite{LISA}.
Potential sources range from massive black hole binaries (MBHBs) \cite{MBHB:sci}, \acp{SBBH} \cite{SBBH:sci}, \acp{EMRI} \cite{EMRI:sci}, 
\Authorfootnote \noindent 
galactic \acp{DWD} \cite{DWD:sci} to \acp{SGWB} \cite{SGWB:sci}.
 
Among these sources, MBHBs are expected to produce the loudest \ac{GW} signals, with a \ac{SNR} of up to several thousand, and can be observed as early as $z \simeq 15 \sim 20$.
MBHBs can merge in gas-rich environments \cite{Barnes:environment}, resulting in a high black hole accretion rate and high star formation rate in the galaxy \cite{Springel:environment, Springel:environment2}.
Both of these factors have the potential to cause strong \ac{EM} radiation, making MBHBs promising targets for multimessenger astronomy.
During the inspiral phase, \ac{EM} emission from MBHBs is predominantly from X-rays emitted by the circumbinary disk \cite{Tang:inspiral}, gradually declines before the merger \cite{Paschalidis:inspiral}, and shifts toward being dominated by ultraviolet radiation as the merger approaches \cite{d'Ascoli:inspiral}.
At the moment of the merger, we anticipate a variety of \ac{EM} emissions, such as a super-Eddington flare from the super-Eddington accretion rate \cite{Armitage:superEddington, Chang:superEddington}, radio emission from the spin flip of the black hole \cite{Merritt:spinflip}, and jets caused by the surrounding magnetic field \cite{Palenzuela:jet, Palenzuela:jet2}.
In addition, with delays ranging from hours to months after coalescence, we may observe highly relativistic jets launched along the spin axis of the black hole \cite{Yuan:jet}.
Even long after the merger, various \ac{EM} afterglows are expected to persist \cite{Shields:XAfterglow, Merritt:XAfterglow, Schnittman:IFAfterglow}.

In recent years, a significant amount of research has focused on studying \ac{GW} and \ac{EM} multimessenger observations involving MBHBs \cite{LISA:multimessenger:prospects, LISA:multimessenger}. 
The primary focus of attention has been the localization capability of \ac{GW} detectors for MBHBs and the anticipated outcomes of joint observations with multiband \ac{EM} facilities.

\ac{GW} observations can provide information about the masses and spins of MBHBs, whereas \ac{EM} observations can provide insights into the environment around massive black holes and reveal the behavior of the accretion disks around the MBHBs, particularly at the later stages of their inspiral evolution.
Multimessenger observations of MBHBs will provide insights into the coevolution of massive black holes, nuclear star clusters, and their host galaxies, shedding light on the time delay between galaxy and MBHB mergers, as well as the physics of active galactic nuclei \cite{DeRosa:AGN}. 
Moreover, the direct measurement of luminosity distance by \ac{GW} analysis and the inference of redshift by \ac{EM} analysis can provide a new measure of the Hubble parameter \cite{Schutz:Hubble, Tamanini:Hubble} and constrain cosmological extra dimensions \cite{Corman:constraindim}. 
By comparing the phases of \ac{GW} and \ac{EM} signals to break the degeneracies of various parameters, the fractional difference in propagation velocity between gravitons and photons can be accurately determined to $10^{-17}$ \cite{Haiman:velocity}.

Despite the promising potential for multimessenger observations of MBHBs, observing the \ac{EM} signal from MBHB mergers poses many challenges primarily because of the short emission timescale.
The typical long distance to the source also results in faint emissions.
Furthermore, distinguishing radiation from background noise can be quite tricky.
The identification of the source as an MBHB would require adequate evidence.

Most of the aforementioned challenges can be addressed by localizing the MBHBs before the merger.
However, the \ac{SNR} of an MBHB accumulates in a highly nonlinear way, with the last hour signal containing up to 99\% of the total \ac{SNR} \cite{Feng:2019wgq} and the sky localization area significantly shrinking as the MBHB merger approaches \cite{LISA:multimessenger}.
Therefore, the localization of MBHBs before the merger raised a new challenge for achieving near real-time speed for the data downlink, as well as for the data analysis.

Because of the relatively short distance from the Earth, TianQin has the potential to enable near real-time data transmission to Earth.
In this study, we analyze the localization of MBHBs before the merger under the assumption that TianQin data can be transmitted in real time.
Because of the longer duration of signals, data analysis of space-borne \ac{GW} detectors can be a lengthy process, taking days or even weeks with the calculation of the likelihood being a major bottleneck.
Algorithms, such as the heterodyned likelihood \cite{Cornish:hetlike, Zackay:hetlike}, reduced order quadratures \cite{Canizares:ROQ}, and multibanding likelihood \cite{Vinciguerra:multibanding, Morisaki:multibanding}, have been developed to expedite the data analysis process.
In this work, we used the heterodyned likelihood algorithm adapted from \texttt{BBHx} \cite{Katz:BBHx}.

This paper is organized as follows: In Section \ref{sec2}, we describe the waveform of MBHBs with the response function of TianQin. In Section \ref{sec3}, we describe the methods used in this work. In Section \ref{sec4}, we apply these methods to the analysis pipeline. In Section \ref{sec5}, we injected three signals to demonstrate the performance of the pipeline. Finally, in Section \ref{sec6}, we discuss the conclusion of this work and the future directions for development.

\section{Waveform of MBHB}\label{sec2}

Throughout this work, we adopt the aligned spin IMRPhenomD \cite{IMRPhenomD1, IMRPhenomD2} waveform for the MBHB signal.
The waveform is described by a set of parameters, that is, $\Theta = \left \{ M_c, \eta, \chi_1, \chi_2, D_L, t_c, \phi_c, \psi, \iota, \lambda, \beta \right \} $.
$M_c$ is the redshifted chirp mass defined as $M_c = (m_1 m_2)^{3/5}/(m_1+m_2)^{1/5} \times (1+z)$.
In this work, we only consider MBHBs with redshifted chirp mass ranging between $10^4 ~M_{\odot}$ and $10^8 ~M_{\odot}$.
$\eta = m_1 m_2/(m_1+m_2)^2$ is the symmetric mass ratio.
Equal mass binaries have a $\eta=0.25$.
We set the lower limit of $\eta$ to $0.05$, which approximates the mass ratio of $1:18$, corresponding to the parameter space where the IMRPhenomD waveforms are reliable.
$\chi_1$ and $\chi_2$ ranging between $-1$ and $1$ are the dimensionless spins of two black holes.
$D_L$ is the luminosity distance, and $t_c$ and $\phi_c$ are the merger time and merge phase, respectively.
$\psi$ is the polarization angle, and $\iota$ is the inclination angle that measures the angle between the angular momentum vector of the binary and the line of sight of the observer.
Finally, we use $\lambda$ and $\beta$ to denote the ecliptic longitude and ecliptic latitude, respectively, of the source location.

Before generating the waveform, we need to determine the frequency range.
The frequency evolution can be calculated using the Newtonian approximation as follows:
\begin{align}\label{foft}
f \left( t \right) = \frac{1}{8\pi} \left( \frac{c^3}{G M_c}\right)^{5/8} \left( \frac{t_c - t}{5}\right)^{-3/8},
\end{align}
If the MBHB merges within the observation period, then the upper limit of the frequency will be determined using the following equation:
\begin{align}
f_{\rm cut} = \frac{1}{5} \frac{c^3}{G M_{\rm tot}},
\end{align}
where $M_{\rm tot}$ is the total mass of the black hole binaries.
We also apply truncation between $10^{-4}$ Hz and $1$ Hz according to the frequency limit of TianQin.
In practice, we adopt \texttt{pyIMRPhenomD} \cite{pyIMRPhenomD} to generate the frequency domain amplitude $\mathcal{A} \left( f \right)$, phase $\Phi \left( f \right)$, and time--frequency relation $t \left( f \right)$.

For space-borne \ac{GW} missions, one of the dominant sources of noise is the laser frequency noise, which can be mitigated through the \ac{TDI} technology \cite{Marsat:TDI, Vallisneri:TDI}.
Throughout this work, we adopt the commonly used orthogonal observables, namely, $A,\ E,$ and noise-insensitive $T$.
These orthogonal observables can be combined through the symmetric \ac{TDI} Michelson channels $X,\ Y,$ and $Z$ \cite{Vallisneri:TDI, Andrzej:TDI}, as follows:
\begin{subequations}
\begin{align}
A&=\frac{1}{\sqrt{2}} \left( Z-X \right), \\
E&=\frac{1}{\sqrt{6}} \left( X-2 Y+Z \right), \\
T&=\frac{1}{\sqrt{3}} \left( X+Y+Z \right),
\end{align}
\end{subequations}
Several \ac{TDI} schemes have been proposed for space-borne \ac{GW} detectors \cite{Cornish:TDI, Shaddock:TDI, Tinto:TDI}. However, these schemes are merely modifications of the first-generation \ac{TDI} and do not significantly alter the response.
As a result, only the first-generation \ac{TDI} will be utilized in this study.
The orthogonal observables can be represented by the basic Doppler observable $\tilde{y}_{slr}$, which represents a laser frequency shift between different types of spacecraft \cite{Marsat:TDI, Vallisneri:TDI, Andrzej:TDI}.

We define $\tilde{A}$ as the Fourier transform of $A$.
Considering the TianQin orbit \cite{TQ:orbit}, we can express the \ac{TDI} response for TianQin of the \ac{TDI} channels in the frequency domain as follows \cite{GWSpace, Lyu:2023, Marsat:TDI}:
\begin{subequations}
\begin{align}
\tilde{A} =& \frac{1}{\sqrt{2}} \left(\mathcal{D}^{2}-1\right) \left[(1+\mathcal{D})\left(\tilde{y}_{31}+\tilde{y}_{13}\right)-\tilde{y}_{23}-\mathcal{D} \tilde{y}_{32}-\tilde{y}_{21}- \right. \nonumber \\
    &\left. \mathcal{D} \tilde{y}_{12}\right] \\
\tilde{E} =& \frac{1}{\sqrt{6}} \left(\mathcal{D}^{2}-1\right) \left[(1-\mathcal{D})\left(\tilde{y}_{13}-\tilde{y}_{31}\right)+(1+2 \mathcal{D})\left(\tilde{y}_{21}-\tilde{y}_{23}\right) \right. \nonumber \\
    &\left. +(2+\mathcal{D})\left(\tilde{y}_{12}-\tilde{y}_{32}\right)\right], \\
\tilde{T} =& \frac{1}{\sqrt{3}} \left(\mathcal{D}^{2}-1\right) (1-\mathcal{D})\left(\tilde{y}_{13}-\tilde{y}_{31}+\tilde{y}_{21}-\tilde{y}_{12}+\tilde{y}_{32}-\tilde{y}_{23}\right) ,
\end{align}
\end{subequations}
where $\mathcal{D} \equiv e^{\frac{{\rm i}2 \pi f L}{c}}$.
As the observable $T$ is insensitive to signals at low frequencies, we concentrate on the $A, E$ channels.

\section{Method}\label{sec3}

\subsection{Bayesian Framework}

In this work, we adopt the Bayesian framework to obtain the posterior distribution of the parameter $\Theta$.
According to Bayes’ theorem, the posterior distribution can be expressed as follows:
\begin{align}
p \left(\Theta \mid d, I \right) = \frac{p \left(d \mid \Theta, I \right) p \left(\Theta \mid I \right)}{p \left(d \mid I \right)},
\end{align}
where $p\left(\Theta \mid d, I \right)$ is the posterior, $p \left(d \mid \Theta, I \right)$ is the likelihood, $p\left(\Theta \mid I \right)$ is the prior, and the $p \left(d \mid I \right)$ is the evidence, with $d$ representing the observed data and $I$ representing the information.

For \ac{GW} data analysis, the likelihood can be expressed as follows:
\begin{align}
  \log \mathcal{L}(\Theta) =& \log p \left(d \mid \Theta, I \right) \nonumber \\
=& -\frac{1}{2}\langle d-h \left(\Theta \right) \mid d-h \left(\Theta\right)\rangle + {\rm const.} \nonumber \\
=& \langle d \mid h \left(\Theta \right)\rangle-\frac{1}{2}\langle h \left(\Theta \right) \mid h \left(\Theta \right)\rangle -\frac{1}{2}\langle d \mid d\rangle + {\rm const.} , 
\end{align}
where $h \left(\Theta \right)$ is the waveform with parameter $\Theta $. $\langle g \mid h \rangle$ represents the inner product between $g$ and $h$ expressed as follows:
\begin{align}
  \langle g \mid h \rangle = 4 \Re \int_0^\infty \frac{\widetilde{g}  \left(f\right) \cdot \widetilde{h}^{*}  \left(f\right) }{S_{n} \left(f\right)} {\rm d}f,
\end{align}
where $S_{n} \left(f\right)$ is the one-sided \ac{PSD} of the noise and $\Re$ is the real component.

The constant is related to the normalization of Bayes’ equation.
If we ignore all constant terms, then the log-likelihood can be simplified as follows:
\begin{align}
  \log \mathcal{L}(\Theta) \propto \langle d \mid h\rangle-\frac{1}{2}\langle h \mid h\rangle.
\end{align}

The analysis of \ac{GW} signals involves high-dimensional parameter spaces.
To efficiently explore the parameter space, stochastic sampling methods, such as \ac{MCMC}, are often used in the \ac{GW} data analysis community.
The \ac{MCMC} method uses random walking of the sampler and encourages moves to the higher posterior region.
In this work, we utilize \texttt{emcee}, a specific realization of the affine invariant ensemble sampler algorithm \cite{emcee}, which enables multiple walkers, representing parameter vectors, to navigate through it by proposing new positions based on the posterior distribution.
The use of multiple walkers enhances the robustness and effectiveness of \ac{MCMC} sampling in complex, high-dimensional parameter spaces \cite{affine:invariance:MCMC}.

Traditionally, \ac{MCMC} methods are only applied to parameter estimation tasks, and the identification of the signal is often treated as a separate scope.
However, in this work, we do not clearly distinguish the \emph{detection} and \emph{measurement} of \ac{GW} signals.
We utilize the \texttt{emcee} to explore the full parameter space, and identify the parameters that maximize likelihood. 
If maximum likelihood exceeding the predefined threshold, we consider the presence of a signal within the data. 
Additionally, this process yields estimations of uncertainties for each parameter.

\subsection{Heterodyned Likelihood}

The overall time spent on the analysis is determined by two factors, that is, the number of samples needed and the average time it takes to generate a single waveform.
In our analysis pipeline, we use the heterodyned likelihood method \cite{Zackay:hetlike, Cornish:hetlike, Katz:BBHx} as part of our \emph{fast estimation} module.

The core idea of the heterodyned likelihood method is to separate the waveform $\widetilde{h}(f)$ into two parts, that is, a rapidly changing component, which is common in the reference waveform $\widetilde{h}_0\left(f\right)$, and a slowly changing component, which indicates the ratio $\widetilde{r} \left(f\right) = \frac{\widetilde{h}\left(f\right)}{\widetilde{h}_0\left(f\right)}$.
Under this decomposition, we can expand the two inner products as follows:
\begin{subequations}
\begin{align}
  \langle d \mid h\rangle & = 4 \Re \int_0^\infty \frac{\widetilde{d} \left(f\right) \cdot \widetilde{h}_{0}^{*} \left(f\right)}{S_{n} \left(f\right)} \times \widetilde{r} \left(f\right) {\rm d}f, \\
  \langle h \mid h\rangle & = 4 \Re \int_0^\infty \frac{ \left| \widetilde{h}_{0} \left(f\right)\right| ^{2}}{S_{n} \left(f\right)} \times \left| \widetilde{r} \left(f\right)\right| ^{2} {\rm d}f,
\end{align}
\end{subequations}
In practice, we first need to obtain a reference waveform $\widetilde{h}_0 \left(f \right)$ with a high posterior.
Then, the rapidly changing components $\widetilde{d} \left(f \right) \widetilde{h}_{0}^{*} \left(f\right) / S_{n} \left(f\right)$ and $\left| \widetilde{h}_{0} \left(f\right)\right| ^{2} / S_{n} \left(f\right)$ only need to be computed once and stored as a precomputed factor.
During sampling, we only need to calculate the waveform ratio $\widetilde{r} \left(f\right)$ on a sparse frequency grid.
In this way, the average time to compute the likelihood can be significantly reduced.

\begin{figure*}[htbp]
  \centering
  \subfigure{\includegraphics[width=0.4\textwidth]{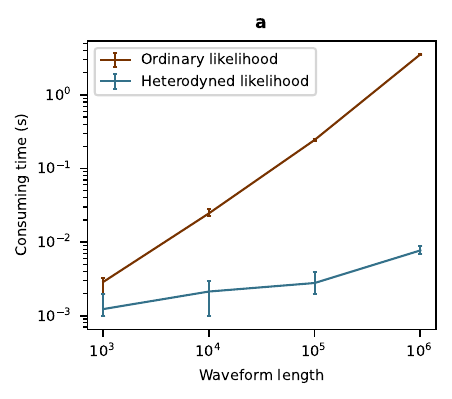}\label{hetliketime}}
  \subfigure{\includegraphics[width=0.4\textwidth]{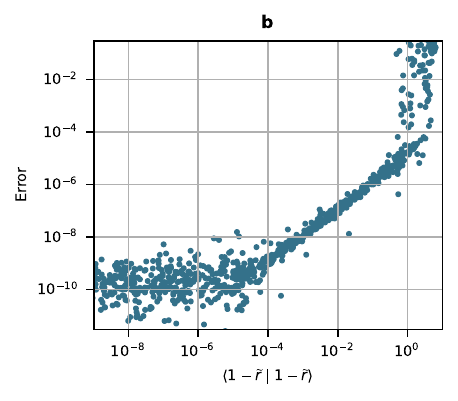}\label{hetlikeerror}}
  \caption{The left panel shows the computing time for waveforms using the heterodyned and ordinary likelihood methods for different waveform lengths. The mean value and 90\% confidence intervals of more than 1,000 waveforms are shown. The right panel shows how the error between heterodyned likelihood and ordinary likelihood is correlated with the deviation of the slow term $\widetilde{r} \left(f \right)$. A strong correlation between the error and $\langle 1-\widetilde{r}| 1-\widetilde{r} \rangle$ can be observed, which indicates that if $\widetilde{r} \left(f \right)$ is indeed slowly changing, then the heterodyned likelihood method can compute the likelihood accurately.}
  \label{hetlike}
\end{figure*}

To quantify the improvement in the efficiency of the heterodyned likelihood method over ordinary likelihood, we generate a total of 4,000 different signals with different waveform lengths ranging from $10^3$ to $10^6$.
We fix $M_c = 3 \times 10^{5} ~M_{\odot}$ and $D_L = 1 ~{\rm Gpc}$ and randomize over all other parameters.
A total of 4,000 different signals with different waveform lengths are generated with both methods. In Figure \ref{hetliketime}, we present the computing time for each case with a single core in CPU Intel Core i7-10700 @ 2.90 GHz, with the lines denoting the mean value and the error bars denoting the 90\% confidence intervals.
Here, the injected waveform is selected as the reference waveform for heterodyned likelihood.
For shorter waveform lengths, the frequency is already sufficiently sparse, and the heterodyned likelihood method does not show a significant advantage against the ordinary likelihood.
However, for longer waveform lengths, the computing time of the heterodyned likelihood increases slower than the ordinary likelihood, leading to a significant improvement in speed.

In addition to efficiency, we can also evaluate the effect of heterodyned likelihood through its accuracy.
If the waveform deviates far from the reference waveform, then the slow term $\widetilde{r} \left(f \right)$ exhibits significant fluctuation in the frequency domain, leading to unacceptable errors of the heterodyned likelihood method.
In Figure \ref{hetlikeerror}, we show the error of heterodyned likelihood.
For this purpose, we select the data shown in Figure \ref{hetliketime} but select reference waveforms with deviations from the injected waveform. The error of the heterodyned likelihood method is defined as the relative error and expressed as follows:
\begin{align}
\mathrm{error} = \left| \frac{\left(\log \mathcal{L}_{\rm ord}\left(\Theta_{\rm inj}\right)-\log \mathcal{L}_{\rm het}\left(\Theta_{\rm inj}\right)\right)}{\log \mathcal{L}_{\rm ord}\left(\Theta_{\rm inj}\right)} \right|. 
\end{align}
The waveform deviation is quantified by $\langle 1-\widetilde{r} \mid 1-\widetilde{r} \rangle$.
Low values of this metric indicate high similarity between new and reference waveforms.
When the waveform bias is small, the likelihood computation error is primarily dominated by numerical errors, approximately at the magnitude of $10^{-9}$.
As the level of waveform deviation surpasses approximately $10^{-4}$, the error from waveform deviation becomes increasingly prominent.
However, even with a waveform deviation of $10^{-2}$, the resulting error is still within an acceptable range of less than $10^{-6}$.

\section{The Analysis Pipeline of TianQin for MBHB}\label{sec4}

For heliocentric orbit missions, such as \ac{LISA}, the data downlink speed/cadence is limited by the availability of deep space networks.
Therefore, providing near real-time data transmission would be relatively challenging and expensive.
By contrast, for geocentric orbit missions, such as TianQin, if inter-satellite communication is enabled and/or multiple ground facilities are available to ensure data downlink, then a reliable and near real-time data downlink becomes feasible.
In an optimistic scenario, TianQin can maintain a data downlink cadence at the order of minutes.
Even in a pessimistic scenario where the near real-time downlink is unavailable all of the time, a data receiving cadence of 2 days can be assumed \footnote{X. Zhang, and Z. Yi,  private communication (2023).}.
In this work, we adopt the assumption of two working modes of the data downlink of TianQin, that is, the \emph{regular} mode, where the full amount of data is available with a latency of 2 days, and the \emph{prompt} mode, where the data can be transferred near real-time.

The \ac{SNR} accumulation of \ac{GW} signals from MBHBs is highly nonlinear.
Therefore, long before the MBHB mergers, the data transmission cadence does not indicate significant differences, and we assume that new data are available every 2 days.
In the first stage, we apply the \emph{search} module that runs every 2 days to routinely check if an upcoming MBHB merger has a sufficiently large \ac{SNR}.
If the \ac{SNR} exceeds the threshold of 8, then we conclude that a signal has been detected.
Otherwise, we conclude that no significant MBHB signal is contained in the data, and the \emph{search} module will retain the original data and rerun the search after receiving a new batch of data.

Once the \ac{SNR} exceeds the threshold, we execute the appropriate module based on the estimated merger time.
The \emph{estimation} module is triggered when the 90\% confidence interval for the estimated merger time is over 1 week after the most recent data reception.
At this point, we still have time to continue with the \emph{regular} data transmission mode and employ ordinary likelihood method for the calculations.
When either the \emph{search} or \emph{estimation} module indicates that the signals may merge within 1 week, we switch to the final stage.
As the MBHB approaches its final merger, we initiate real-time data transmission in the final stage.
Employing the \emph{fast estimation} module, we regularly perform parameter estimation and update the posterior distribution of the full parameter set.
As the merger of the MBHB approaches, the uncertainty in the estimated parameters quickly diminishes \cite{LISA:multimessenger}.
This iterative process continues until we estimate that the MBHB has merged.
At this point, the signal associated with the estimated parameters is removed from the data, and the \emph{search} module is rerun to scour the data for the next potential signal unhindered by the previously estimated one.
This cycle of searching, estimating, and subtracting continues until TianQin suspends its observation.

\begin{figure*}[htbp]
  \centering
  \includegraphics[width=0.8\textwidth]{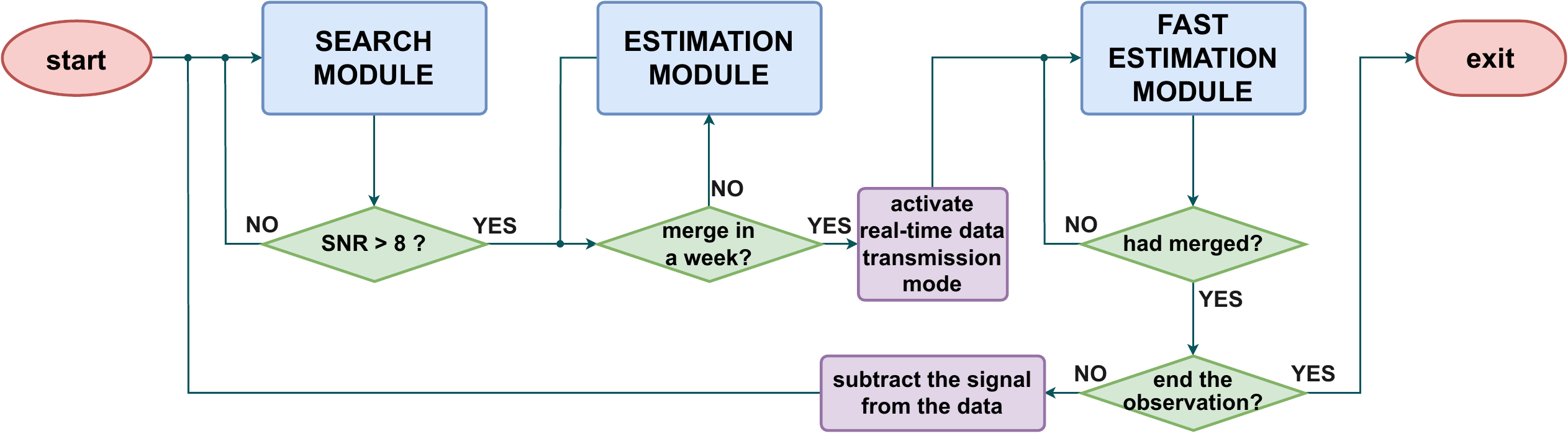}
  \caption{Flowchart of the designed pipeline. The data analysis can be separated into three modules, namely, the \emph{search}, \emph{estimation}, and \emph{fast estimation} modules. The \emph{search} module routinely searches for indicators of upcoming MBHB mergers, the \emph{estimation} module works if the MBHB merges after 1 week, and the \emph{fast estimation} module works when a real-time downlink is enabled.}
  \label{pipeline}
\end{figure*}

\begin{table}[H]
\centering
\footnotesize
\begin{threeparttable}\caption{Bounds for all parameters}\label{prior}
\doublerulesep 0.1pt \tabcolsep 13pt 
\begin{tabular}{ccc}
\toprule
\textrm{Parameter} & \textrm{Lower Bound} & \textrm{Upper Bound}\\
\midrule
$\log \left(M_c/M_{\odot}\right)$ & $\log \left(10^4\right)$ & $\log \left(10^8\right)$\\
$\eta$ & $0.05$ & $0.25$\\
$\chi_1$ & $-1.0$ & $1.0$\\
$\chi_2$ & $-1.0$ & $1.0$\\
$D_L/{\rm Gpc}$ & $0.01$ & $230$\\
$t_c/{\rm day}$ & $0$ & $90$\\
$\phi_c/{\rm rad}$ & $0$ & $2\pi$\\
$\psi/{\rm rad}$ & $0$ & $\pi$\\
$\cos \iota$ & $-1$ & $1$\\
$\lambda/{\rm rad}$ & $0$ & $2\pi$\\
$\sin \beta$ & $-1$ & $1$\\
\bottomrule
\end{tabular}
\end{threeparttable}
\end{table}

In Figure \ref{pipeline}, we use a flowchart to summarize the logic of the entire pipeline.
The same prior is adopted throughout all modules.
For parameters that vary in orders of magnitude, such as $M_c$, a log-uniform prior is adopted.
For $D_L$ less than the cutoff $D_{\rm max}=230 ~{\rm Gpc}$ (or $z\approx 20$), we adopt a uniform prior in comoving volume, as follows:
\begin{align}
  \log{p \left(D_L \mid I \right)} \propto 2\log{D_L}, \quad {\rm for\ }D_L<D_{\rm max}
\end{align}
For the other parameters, we adopt a uniform prior either for the parameter \emph{per se} or over the sphere.
The bounds for all parameters are shown in Table \ref{prior}.
Note that because TianQin adopts the “3 months on + 3 months off” working scheme and most of the \ac{SNR} is obtained just before the merger, we focus on the scenario that the MBHB merges within the continuous 3-month period. Thus, the upper limit of the merger time is set to 90 days.

For all \ac{MCMC} processes, we have chosen the logarithmically uniform initial point for $M_c$.
For sky positioning and $\iota$, we select the uniform initial points on the sphere.
In addition, we determined that fixing the initial point of $t_c$ to the upper limit (3 months) enhances the sampling efficiency.
For the initial points of the other parameters, we randomly select points from a uniform distribution.

\subsection{Search Module}

Using the \emph{search} module, we perform an MCMC-based analysis with 300,000 steps and 24 walkers.
Note that, in this stage, we evaluate likelihood using the ordinary likelihood, which does not significantly increase the computational burden because the frequency evolution of the binary black hole during its inspiral phase is considerably slow.
We only need to perform a calculation using limited frequency points.
In practice, this module takes approximately 10 h on 24 cores in CPU Intel Xeon Silver 4210R @ 2.40 GHz.
After the \emph{search} module, the \ac{MCMC} algorithm yields the maximum posterior probability and posterior distributions for all parameters.
We calculate the \ac{SNR} of the signal based on the parameter set associated with the maximum posterior.

\subsection{Estimation Module}

The \emph{estimation} module is similar to the \emph{search} module, with the only difference being focusing attention on parameter estimation.
The initial points are drawn from the 90\% confidence interval of the previous analysis instead of the prior distribution as in the \emph{search} module.
This change can reduce the time spent on burn-in.
Experiments indicate that performing an MCMC-based analysis with 24 walkers and 100,000 steps is sufficient.
In practice, this module will take not more than $5$ h, using the same hardware as the \emph{search} module.

\subsection{Fast Estimation Module}

Once a signal is detected and estimated to merge within 1 week, the \emph{fast estimation} module is utilized to analyze the data.
The \emph{fast estimation} module consists of two parts, that is, a quick optimization using the Nelder--Mead (NM) algorithm \cite{NM} and an analysis using the heterodyned likelihood-based \ac{MCMC} algorithm.

Because of the small inherent biases often found in the inferred maximum likelihood parameters, previous analysis results may exhibit a significantly low likelihood when confronted with new data.
Consequently, directly utilizing the previous analysis results as the reference waveform for the heterodyned likelihood in the subsequent analysis is infeasible.
Therefore, this work employs the NM method from \texttt{scipy.optimize} to obtain a reference waveform for the heterodyned likelihood.
The NM method, a commonly used optimization technique, iteratively optimizes a nonlinear objective function without requiring gradient information, working by defining a simplex and iteratively modifying its vertices to explore the search space and converge toward the optimal solution.

We initiate the NM process using the maximum posterior estimation value from the previous analysis as a starting point.
With a maximum iteration limit of 300, we aim to identify the parameters that maximize the ordinary likelihood within the 90\% confidence interval established in the previous analysis.
In terms of speed, NM significantly outperforms \ac{MCMC}, completing the task in less than 5 min.

To assess the performance of NM against other point estimation algorithms, we compare it with differential evolution \cite{DE} and particle swarm optimization \cite{PSO} from the \texttt{scikit-opt} library.
In our pipeline, all the aforementioned algorithms exhibit comparable computational accuracy, yielding reliable results with deviations from the injected signal that are less than $10^{-2}$. However, NM emerges as the fastest among them.

Once we obtain the reference parameters from NM, we proceed to apply the heterodyned likelihood-based \ac{MCMC} algorithm.
Heterodyned likelihood effectively boosts the calculation speed of the likelihood function with only negligible errors, enabling us to obtain the results in approximately 40 min, leveraging 24 walkers and 100,000 steps in the \ac{MCMC} process.

\section{Result}\label{sec5}

\begin{figure}[H]
  \centering
  \includegraphics[width=0.4\textwidth]{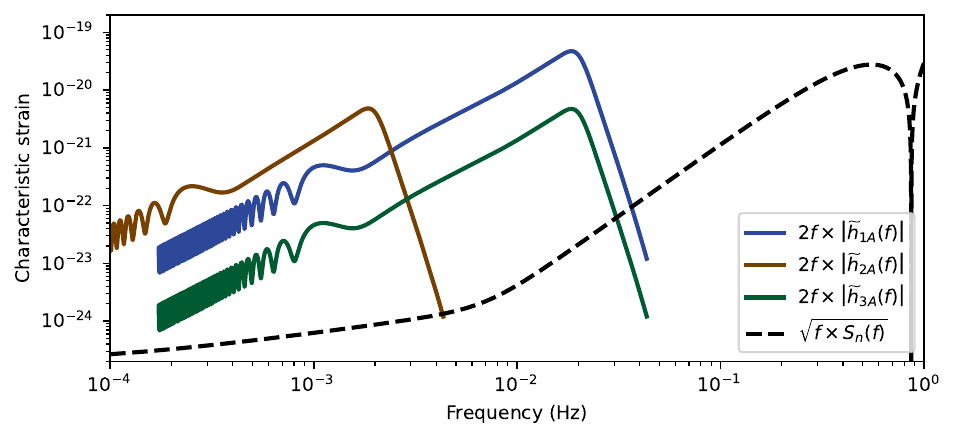}
  \caption{Each datum contains an MBHB signal. This figure shows the A channel waveforms for the three injected sources and the sensitivity curves in the representation of the frequency domain characteristic strain.}
  \label{signal}
\end{figure}

To test the performance of the pipeline, we perform a near real-time analysis of the simulated data.
For the Gaussian noise, we generate signals according to the one-sided \ac{PSD} of TianQin \cite{TianQin}.
In Table \ref{inj_para}, we summarize the parameters of the three simulated events.
The first event represents the ideal case for TianQin, the second event has a heavier system, and the third event has a lower \ac{SNR}.
For all cases, binaries will merge 2 months after starting the TianQin observation.
In Figure \ref{signal}, we present the characteristic strain of all three injected waveforms for the A channel.
The noise amplitude $\sqrt{f S_n(f)}$ is denoted by a black dashed line, where $S_n(f)$ is the one-sided \ac{PSD}.

\begin{table}[H]
\centering
\footnotesize
\begin{threeparttable}\caption{Parameters of three injected sources}\label{inj_para}
\doublerulesep 0.1pt \tabcolsep 13pt 
\begin{tabular}{cccc}
\toprule
\textrm{Parameter} &
\textrm{Source 1} &
\textrm{Source 2} &
\textrm{Source 3} \\
\midrule
$M_c/M_{\odot}$ & $3 \times 10^{5}$ & $3 \times 10^{6}$ & $3 \times 10^{5}$ \\
$\eta$ & $0.15$ & $0.15$ & $0.15$ \\
$\chi_1$ & $0.5$ & $0.5$ & $0.5$ \\
$\chi_2$ & $0.7$ & $0.7$ & $0.7$ \\
$D_L/{\rm Gpc}$ & $1$ & $1$ & $10$ \\
$t_c/{\rm day}$ & $60$ & $60$ & $60$ \\
$\phi_c/{\rm rad}$ & $\pi/2$ & $\pi/2$ & $\pi/2$ \\
$\psi/{\rm rad}$ & $\pi/4$ & $\pi/4$ & $\pi/4$ \\
$\iota/{\rm rad}$ & $\pi/8$ & $\pi/8$ & $\pi/8$ \\
$\lambda/{\rm rad}$ & $\pi$ & $\pi$ & $\pi$ \\
$\beta/{\rm rad}$ & $\pi/3$ & $\pi/3$ & $\pi/3$ \\
${\rm SNR}$ & $4242$ & $4067$ & $425$\\
\bottomrule
\end{tabular}
\end{threeparttable}
\end{table}

\subsection{SNR Accumulation}

\begin{figure}[H]
  \centering
  \includegraphics[width=0.4\textwidth]{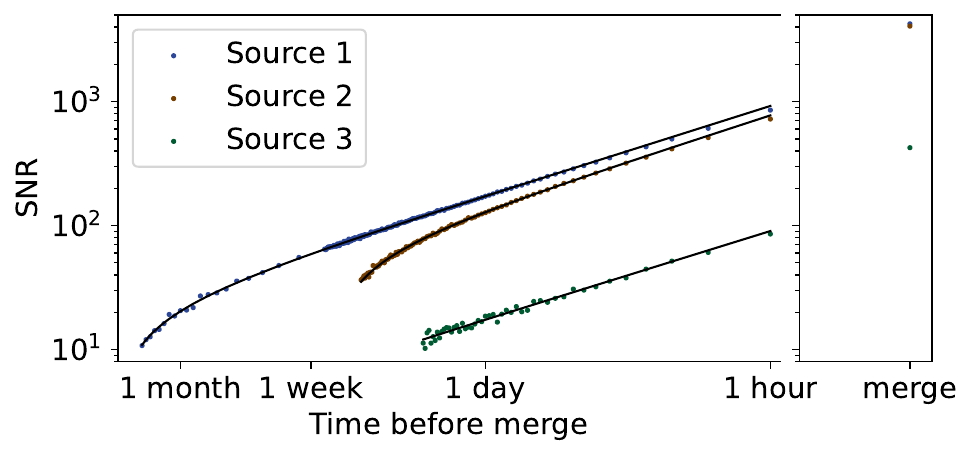}
  \caption{\ac{SNR} accumulation curves of the three injected sources. The dots denote the \ac{SNR} recovered at different times, and the black line denotes the result of using the relationship (\ref{snr}) during fitting. The horizontal axis shows the time before the merger, and the vertical axis shows the \ac{SNR}; both axes are shown in logarithmic scale.}
  \label{SNR}
\end{figure}

The \ac{SNR} of the \ac{GW} signal plays a crucial role in the parameter estimation of MBHB systems.
The uncertainties of many parameters, for example, luminosity distance, are inversely proportional to the \ac{SNR}.
We first provide a quantitative demonstration of the nonlinear accumulation of \ac{SNR} over the observation time.
Figure \ref{SNR} shows the \ac{SNR} accumulation of three different sources at various observation times.
In the right panel, we illustrate the total \ac{SNR} of the complete data.
We observe that the last hour of data contributes significantly to the \ac{SNR} with a huge jump between lines across the panels.
Both Sources 1 and 2 have relatively short distances; thus, they share a similar total \ac{SNR}.
Source 1 took 14 days to reach the \ac{SNR} threshold of 8.
However, Source 2 is more massive and has a lower \ac{GW} frequency.
Therefore, the system only enters the observation band of TianQin 6 days before its merger.
Subsequently, because of the rapid accumulation of \ac{SNR} near the merger, our next analysis (4 days before the merger) detected Source 2 with an \ac{SNR} of approximately 30.
For Source 3, because of its relatively low \ac{SNR}, it could not surpass the SNR threshold of 8 until 2 days before the merger.

Because of the TianQin low-frequency ($f \le 6 ~mHz$) \ac{TDI} sensitivity curve that resembles a straight line, the \ac{SNR} of the MBHB inspiral increases based on the following power law:
\begin{align}\label{snr}
{\rm SNR} \propto \left(\frac{t}{t_c - t}\right)^{1/2}.
\end{align}
This relationship formula can also be calculated using the frequency--time relationship(\ref{foft}), post-Newtonian approximation \ac{GW} waveform.
We also observe that the inclusion of Gaussian noise introduces fluctuations.

\subsection{Sky Localization}

The main motivation for this work is to localize the massive black holes during their merger, enhancing the possibility of successful multimessenger observations.
Thus, obtaining accurate and timely sky localization capabilities is paramount.
We demonstrate the evolution of estimated sky localization uncertainties over different times for the three injected signals.
To gain a better understanding of their performance, some typical \acp{FOV} of flagship \ac{EM} telescopes were listed for comparison.

\begin{figure*}[htbp]
  \centering
  \includegraphics[width=0.8\textwidth]{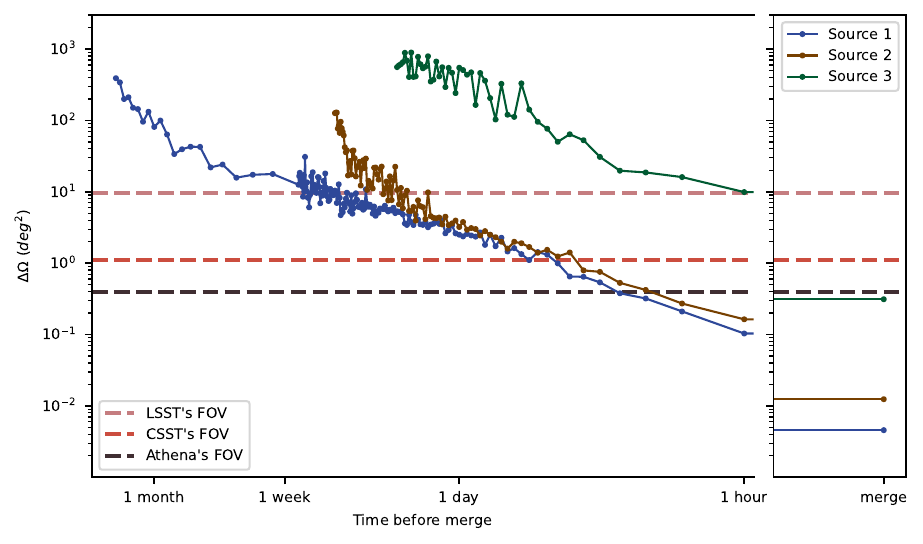}
  \caption{Time evolution of the $90\%$ confidence interval of sky localization uncertainty $\Delta \Omega$ of the three injected sources. For reference, the fields of view of LSST  ($\sim$ 9.6 deg$^2$), CSST ($\sim$ 1.1 deg$^2$), and Athena ($\sim$ 0.4 deg$^2$) are denoted by dashed lines.}
  \label{localizationError}
\end{figure*}

In Fig \ref{localizationError}, we show the time evolution of the 90\% confidence interval of sky localization error $\Delta \Omega$ of the three injected sources. 
For reference, the \ac{FOV} of three telescopes is denoted by dashed lines.
The fluctuations in the localization arise from the fluctuations in the accumulation of \ac{SNR} and the stochastic sampling process.
For comparison, we show the \ac{FOV} of three representative telescopes, that is, the Vera Rubin Telescope (LSST) \cite{LSST, LSST2}, the Chinese Space Station Telescope (CSST) \cite{CSST}, and the Advanced Telescope for High-Energy Astrophysics (Athena) \cite{Athena}.
LSST is a wide-field ground-based system with a 9.6 deg$^2$ \ac{FOV}, designed to study various objects in the universe with advanced technology. CSST is a space telescope with a 1.1 deg$^2$ \ac{FOV}, designed and developed in China and planned to be launched and assembled in orbit as part of the Chinese space station project. Athena is a high-energy astrophysics observatory designed by the ESA with a 0.4 deg$^2$ \ac{FOV}, designed to study celestial objects emitting X-rays.
All of these telescopes are expected to operate during the observation period of TianQin and can perform multimessenger observations of MBHBs.

For Source 1 (2), TianQin can successfully localize it within the \ac{FOV} of LSST approximately 1 week (2 days) before the merger.
Because they share comparable \acp{SNR}, the evolution of the localization uncertainties of Sources 1 and 2 are similar 1 day before the final merger.
In the final hours, both events can be localized in a smaller area than the \ac{FOV} of CSST and Athena.
For Source 3, because it is weaker and only detectable 2 days before the merger, the localization uncertainty is significantly larger than that of stronger signals.
The localization uncertainty only narrows down to the area comparable to the \ac{FOV} of LSST in the final hour before the merger and approximately converges to the \ac{FOV} of Athena at the time of the merger.
The localization uncertainty also exhibits strong fluctuations, which are mainly attributed to the randomness of noise and the random selection of MCMC random number seeds.

With real-time data transmission and analysis capabilities, the localization of MBHBs can be achieved reliably several hours before their final merger.
The precision of this localization is high such that a single snapshot from wide-field telescopes, such as LSST, can potentially cover the entire region of interest, which significantly simplifies the \ac{EM} follow-up observation strategy, as discussed in detail in \cite{Chan:2015bma}.
By contrast, if only the regular data transmission mode is available, then there may be a delay of up to 2 days before receiving a new batch of data.
Under such circumstances, only the stronger sources can be localized to a level comparable to the \ac{FOV} of LSST.
This delay can lead to the loss of valuable information about the merger, as EM telescopes may miss the target at a critical moment.

\begin{figure*}[htbp]
  \centering
  \subfigure{\includegraphics[width=0.4\textwidth]{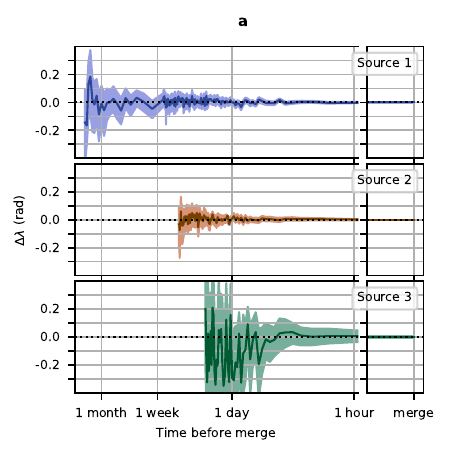}\label{lambda}}
  \subfigure{\includegraphics[width=0.4\textwidth]{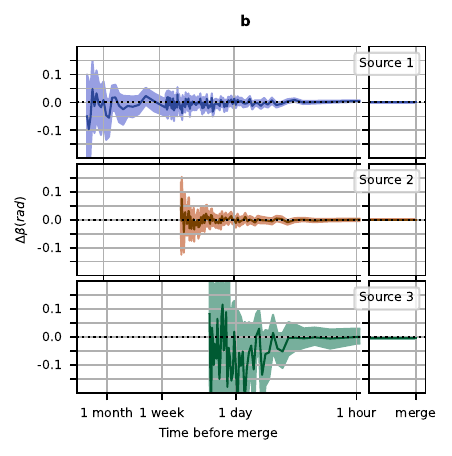}\label{beta}}
  \caption{Evolution of the estimation error of the ecliptic longitude (left panel) and ecliptic latitude (right panel) of the three injected sources. The shaded region denotes the 90\% confidence interval, and the solid line denotes the error of the mean value.}
  \label{localization}
\end{figure*}

In Fig \ref{localization}, we present the evolution of the estimation error of the ecliptic latitude and ecliptic longitude.
Compared with ecliptic longitude, ecliptic latitude can be slightly better constrained by TianQin because of its configuration.
The measurement of the longitude and latitude of the source by the space GW detector relies on different mechanisms. Thus, we do not anticipate them to be precisely identical.
The localization of the source by the space GW detector mainly comes from the motion of the detector (low-frequency) and Doppler modulation (high-frequency).
The motion of the detector has a better capability to limit the longitude, whereas Doppler modulation has a better capability to limit the latitude.
We remark that the injected value lies consistently within the 90\% confidence intervals throughout the entire process, indicating that our near real-time analysis can present reliable sky localization.

\subsection{Other Parameters Constraint}

Finally, we discuss TianQin's capability to constrain the other parameters of the three sources, especially the luminosity distance and merger time.
The accurate measurement of the luminosity distance plays a significant role in reducing the number of potential host galaxies.
Furthermore, a precise estimation of the merger time is crucial for coordinated multimessenger observations.

\begin{figure*}[htbp]
  \centering
  \subfigure{\includegraphics[width=0.4\textwidth]{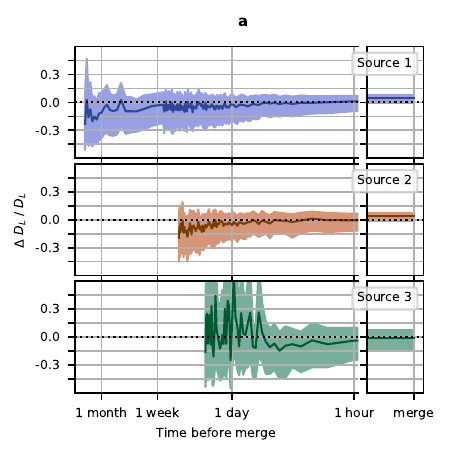}\label{dl}}
  \subfigure{\includegraphics[width=0.4\textwidth]{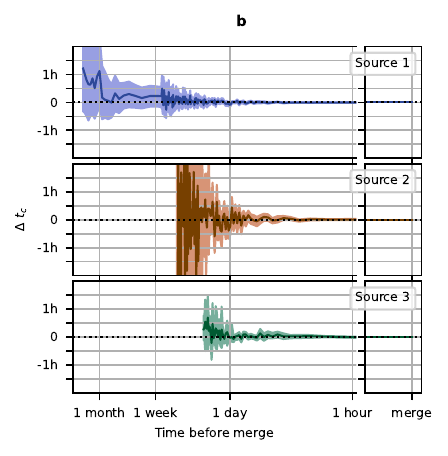}\label{tc}}
  \caption{Similar to Figure \ref{localization}, but for the relative errors of luminosity distance (left panel) and the errors of merger time in units of seconds (right panel).}
  \label{DLtc}
\end{figure*}

Localizing sources in three-dimensional space requires not only the ecliptic longitude and ecliptic latitude but also the accurate determination of the luminosity distance.
Notably, a given sky area could potentially encompass multiple galaxies.
When aiming to utilize \ac{GW} observation of MBHBs to infer the cosmological parameters, the successful identification of the host galaxy would be important \cite{Zhu:2021aat,Zhu:2021bpp}.
Figure \ref{dl} illustrates the evolution of the uncertainties of the luminosity distance of the three sources.
For the three sources, the final uncertainties of the luminosity distance are 7.9\%, 8.5\%, and 22.0\%.
Although the \acp{SNR} of the three signals is high, the degeneracy between the luminosity distance and the inclination angle limits the precision of the measurement.
Future inclusion, such as the inclusion of higher modes \cite{HM:Lionel} or the employment of a network of multiple detectors \cite{network:Kevin}, could potentially alleviate this degeneracy and enhance the accuracy of luminosity distance measurements.

In this study, we choose to activate the real-time data analysis when the MBHB is anticipated to merge within 1 week.
Consequently, the precise and reliable measurement of the merger time is critical to this work.
In Figure \ref{tc}, we exhibit the evolution of the estimated merger time uncertainties.
Upon initial detection of the signal, the merger time estimation for Sources 1 and 3 may have an error of several hours.
By contrast, for Source 2, the uncertainty in the merger time could exceed 1 day because of the insufficient frequency evolution.
However, the merger time uncertainty of Source 2 can be significantly reduced over a timescale of hours.
Within 1 day before the merger, the uncertainty narrows down to 10 min for Source 1 and 1 h for Sources 2 and 3.
When considering data including the merger, the merger time estimation uncertainties of the three sources further decrease to 0.7, 6.6, and 23.2 s. 
In all cases, the merger forecast remains sufficiently accurate to ensure timely real-time data transmission.

\begin{table*}[htbp]
\centering
\footnotesize
\caption{Bias and 90\% confidence intervals recovered from the final analysis after the merger. The bias represents the deviation between the estimated mean value and the injected value. The columns represent different sources, and the rows represent different parameters.}
\label{rec_para}
\renewcommand{\arraystretch}{1.5}
\tabcolsep 18pt 
\begin{tabular}{cccc}
\toprule
\textrm{Parameter} & 
\textrm{Source 1} &
\textrm{Source 2} &
\textrm{Source 3} \\
\midrule
$\Delta M_c/M_{\odot}$ & 
$+3.42 ^{+4.24}_{-4.18}$ & 
$-26.98 ^{+486.08}_{-502.31}$ & 
$+43.21 ^{+44.94}_{-47.11}$ \\
$\Delta \eta$ & 
$\left(-4.34 ^{+9.20} _{-9.32}\right) \times 10^{-5}$ & 
$\left(-0.42 ^{+6.00}_{-6.12}\right) \times 10^{-5}$ & 
$\left(-0.35 ^{+1.13}_{-1.04}\right) \times 10^{-3}$ \\
$\Delta \chi_1$ & 
$\left(+2.75 ^{+6.12}_{-6.14}\right) \times 10^{-4}$ & 
$\left(-1.83 ^{+5.53}_{-5.87}\right) \times 10^{-4}$ & 
$\left(-0.19 ^{+0.94}_{-1.30}\right) \times 10^{-2}$ \\
$\Delta \chi_2$ & 
$\left(-1.53 ^{+2.50}_{-2.51}\right) \times 10^{-3}$ & 
$\left(+0.89 ^{+1.85}_{-1.87}\right) \times 10^{-3}$ & 
$\left(+1.20 ^{+6.43}_{-4.49}\right) \times 10^{-2}$ \\
$\Delta D_L/{\rm Gpc}$ & 
$\left(+4.62 ^{+2.94}_{-4.98}\right) \times 10^{-2}$ & 
$\left(+4.32 ^{+3.21}_{-5.28}\right) \times 10^{-2}$ & 
$-0.14 ^{+0.86}_{-1.34}$ \\
$\Delta t_c/{\rm s}$ & 
$+0.20 ^{+0.36}_{-0.35}$ & 
$-1.66 ^{+3.33}_{-3.30}$ & 
$-3.63 ^{+9.40}_{-13.75}$ \\
$\Delta \phi_c/{\rm rad}$ & 
$-0.28 ^{+0.86}_{-1.12}$ & 
$-0.49 ^{+1.10}_{-0.92}$ & 
$-0.53 ^{+1.15}_{-0.91}$ \\
$\Delta \psi/{\rm rad}$ & 
$+0.41 ^{+1.06}_{-1.04}$ & 
$+0.48 ^{+0.93}_{-1.10}$ & 
$+0.30 ^{+1.10}_{-0.94}$ \\
$\Delta \iota/{\rm rad}$ & 
$-0.14 ^{+0.15}_{-0.16}$ & 
$-0.13 ^{+0.15}_{-0.17}$ & 
$-0.04 ^{+0.25}_{-0.27}$ \\
$\Delta \lambda/{\rm rad}$ & 
$\left(+1.92 ^{+9.51}_{-8.01}\right) \times 10^{-4}$ & 
$\left(-0.64 ^{+1.44}_{-1.40}\right) \times 10^{-3}$ & 
$\left(-0.78 ^{+8.92}_{-7.60}\right) \times 10^{-3}$ \\
$\Delta \beta/{\rm rad}$ & 
$\left(+3.17 ^{+5.84}_{-5.85}\right) \times 10^{-4}$ & 
$\left(+6.67 ^{+9.83}_{-9.76}\right) \times 10^{-4}$ & 
$\left(-3.67 ^{+4.28}_{-4.21}\right) \times 10^{-3}$ \\
\bottomrule
\end{tabular}
\end{table*}

In Table \ref{rec_para}, we present the estimated uncertainties of all 11 parameters of the MBHBs, using all data including the merger and ringdown phase.
In terms of relative uncertainty, the redshifted chirp mass is the most precisely constrained parameter.
For Source 1, the confidence interval of the chirp mass is confined to approximately $3 \times 10^{-5}$ around the true value.
For Sources 2 and 3, the confidence interval of the chirp mass can be narrowed down to approximately $3 \times 10^{-4}$.
In addition, the merger phase and polarization angle are nearly entirely unconstrained. 
The uncertainty of the inclination angle is also substantial, which hinders the precise constraining of the luminosity distance.

\section{Discussion}\label{sec6}

The multimessenger observation of MBHBs has profound scientific significance, but real-time data transmission and analysis capabilities of \ac{GW} observation are essential for conducting multimessenger observations.
The geocentric orbit of the TianQin mission facilitates real-time data transmission, enabling prompt scientific insights.
We consider two modes of data transmission, namely, \emph{regular} and \emph{prompt}.
The regular mode operates at a steady pace, suitable for routine observations, whereas the prompt mode is activated in response to a predicted MBHB merger within 1 week.
In this study, we focus on the TianQin mission and devise a rapid analysis pipeline to address the challenges.
We used a series of methods, such as heterodyned likelihood, to accelerate the calculation speed, which improved our capability to capture the key scientific moment, that is, the merger of MBHBs, in this pipeline.

We inject three simulated signals into Gaussian noise and test the performance of our data analysis pipelines across different operating modes.
With the approaching merger, each analysis can be completed in just 40 min.
For sources with higher \ac{SNR}, the localization region can be narrowed down to within 1 deg$^2$ on the final day before the merger.
For sources with lower \ac{SNR}, the same level of localization uncertainty can be reached with post-merger data.
The results indicate the necessity of real-time transmission.
Without this, accurate sky positioning of the majority of MBHBs before the merger would be significantly more challenging.
After each analysis is completed, TianQin can update the early warning and localization of the source in real time, notify the \ac{EM} telescope, and conduct joint multimessenger observations.
In addition, we demonstrated the capability of TianQin to constrain the other parameters of MBHBs.

As a preliminary exploration, our work suffers from potential caveats.
For example, we simulate and analyze the noise under Gaussian and stationary assumptions, which may not accurately reflect real-world scenarios.
We might also need to update the \ac{PSD} when a more realistic prediction is available.
For the injected signal, the IMRPhenomD waveform model used in this work excludes higher-order modes, which can potentially improve parameter estimation and sampling efficiency by breaking degeneracies, such as luminosity distance--inclination angle degeneracy.
Future work will explore the use of higher-order modes for faster and more accurate analysis.
Furthermore, we ignore the impact of all other sources, such as galactic binaries, or practical challenges, such as data gaps.
Finally, this work has completed a series of analyses using only a few cores.
If we have a large number of CPUs, then we can lower the SNR threshold, increase the search sensitivity, and use a large number of CPUs to determine whether the candidate is a true signal.
We leave the inclusion and treatment of more realistic issues for future exploration.

\Acknowledgements{This work has been supported by Guangdong Major Project of Basic and Applied Basic Research (Grant No. 2019B030302001), and the Natural Science Foundation of China (Grants No. 12173104, No. 12261131504).
We would like to thank Jian-Dong Zhang, Shuai Liu, Xue-Ting Zhang, Lu Wang and Han Wang for helpful comments. 
We are indebted to Xue-Feng Zhang and Zhao-Xiang Yi for constructive discussions on the data downlink of TianQin. }

\InterestConflict{The authors declare that they have no conflict of interest.}



\bibliographystyle{scpma}
\bibliography{sn-bibliography}

\end{multicols}
\end{document}